# COVID-19 Diagnosis: ULGFBP-ResNet51 approach on the CT and the Chest X-ray Images Classification

Vida Esmaeili[1], Mahmood Mohassel Feghhi[*,1], Seyed Omid Shahdi[2]

[1] Faculty of Electrical and Computer Engineering, University of Tabriz, Tabriz, Iran

[2] Department of Electrical Engineering, Qazvin Branch, Islamic Azad University, Qazvin, Iran

*Abstract*-- **The contagious and pandemic COVID-19 disease is currently considered as the main health concern and posed widespread panic across human-beings. It affects the human respiratory tract and lungs intensely. So that it has imposed significant threats for premature death. Although, its early diagnosis can play a vital role in revival phase, the radiography tests with the manual intervention are a time-consuming process. Time is also limited for such manual inspecting of numerous patients in the hospitals. Thus, the necessity of automatic diagnosis on the chest X-ray or the CT images with a high efficient performance is urgent. Toward this end, we propose a novel method, named as the ULGFBP-ResNet51 to tackle with the COVID-19 diagnosis in the images. In fact, this method includes Uniform Local Binary Pattern (ULBP), Gabor Filter (GF), and ResNet51. According to our results, this method could offer superior performance in comparison with the other methods, and attain maximum accuracy.**

*Keywords*-- **COVID-19 disease diagnosis, ULGFBP-ResNet51, CT dataset, X-ray dataset, deep learning.**

## I. INTRODUCTION

Since early months of 2020, coronavirus disease (COVID-19), which is considerably contagious has permeated through the globe [1, 2]. It has imposed significant and unprecedented sufferings and threats for premature death [2]. Unequivocally, it is now regarded as the most deadly and dangerous disease that makes severe panic to the crowd [3]. The well-known reason for death of this pandemic is obstacles in oxygen intake due to inflammation lung, filled air sacs with discharge and fluid [3]. Early identification of the COVID can not only reduce death rate sharply, but also most prone to faster recovery phase [1].

For the first time in the December of 2019, the sick persons infected with COVID-19 were identified in Wuhan, China [4]. Often, the patients develop a dry cough, fever, shortness of breath, weariness, sore throat, pains, runny nose, body aches, and diarrhoea symptoms. High fever and dry cough are its core symptoms [3]. Its symptoms are similar to pneumonia and influenza-A that affect the human respiratory tract and lungs [1, 5]. Since the separation of infection between COVID-19 and bacterial pneumonia is not an easy task, the automatic feature extraction from images can help to diagnose the disease [6]. The difference is that lung lesions in COVID-19 patients are higher than pneumonia and influenza diseases [7]. In fact, COVID-19 damages the lungs intensely. The virus causes the demise of most persons who have chronic diseases (for instance, diabetes) [8].

The viability of this virus in the air is expected to be for almost three hours [3]. It can travel through the patient's cough or sneeze droplets from person to person in close contact. It can even contaminate humans with eating food in infected copper, plastic, and stainless steel dishes. It should be mentioned out the COVID-19 can be live in aforementioned utensils for several hours [3].

Several diagnostic tasks such as viral throat swab testing, blood, and serologic tests are conducted for this disease. Also, Reverse Transcriptase-Polymerase Chain Reaction (RT-PCR) is a yardstick from Nasopharyngeal Swabs (NS) and Or-pharyngeal Swabs (OS) samples. Nevertheless, these recognition measures do not only require manual intervention but also are time-consuming processes [2, 9]. Therefore, using the X-ray or Computed Tomography (CT) data is more preferable [10, 11]. These scanning images conspicuously indicate COVID-19 viral infections with higher confidence. Although, these medical imaging modalities are available and

---

[*] Corresponding author: mohasselfeghhi@tabrizu.ac.ir

[1]Address: 29 Bahman Blvd., University of Tabriz, Tabriz, Iran 5166616471. Tel.: +98-4133393635; fax: +98-4133344272.

Email addresses: v.esmaeili@tabrizu.ac.ir (Vida Esmaeili), mohasselfeghhi@tabrizu.ac.ir (Mahmood Mohassel Feghhi), shahdi@qiau.ac.ir (Seyed Omid Shahdi)

inexpensive, they are not rich in resolution. X-ray image is obtained much faster than its CT counterpart. Meanwhile, CT provides large data appropriate for deep learning methods [12-14].

Due to the lack of certainty in clinical methods, it is necessary for disease diagnosis to accompany with computer-aided high-end accuracy[15]. In health care system the time is always precious since it is limited for numerous patients in hospitals; hence manual diagnosis procedure could be painstaking. To save as many lives, image processing and understanding techniques could play a main role at recognition task at hand. Nonetheless, it should be performed quickly and highly efficiently to be fully fruitful and as a result avoid human errors.

To this end, computer vision, machine learning, and deep learning approaches are put into trial for the efficient diagnosis of radiology images. In this paper, we categorize them into hand-engineered, deep-learning, and mixture methods. The hand-engineered approach extracts features using some predefined patterns in images. It includes Gabor Filter (GF), Local Binary Pattern (LBP), Histograms of Oriented Gradients (HOG), and etc. [6, 16-23].

Deep-learning approach (e.g. Convolutional Neural Network) learns from features extracted from raw images [24-28]. The advantages and disadvantages of some available hand-engineered and deep-learning methods for COVID-19 identification are presented in Table 1. The mixture method makes use of both hand-engineered and deep-learning methods [3, 8, 29-31]. In fact, it combines them in the favor of reaching more benefits.

In the manuscript, we adopt a novel mixture approach named ULGFBP-ResNet51 in order to identify COVID-19 on medical images. In fact, this method includes Uniform LBP (ULBP), GF, and ResNet51. According to each of these comprising method characteristics, this method will outperform in various aspects. The design diagram of our proposed method is illustrated in Fig. 1. The main contributions of this paper are as follows:

- Proposing a novel mixture method named ULGFBP-ResNet51 for COVID-19 diagnosis

- Using the ULGFBP map as the network input rather than original data. Unlike LBP, ULBP is not sensitive to some conditions such as rotation and illumination. GF is also robust to features in multi-scale and direction. Therefore, the outcome map using these methods tends to have better performance on classification.

- Achieving high accurate results for COVID-19 identification in comparison with other related works.

The remaining parts of the manuscript are formed as follows:

Related literature is expressed in the next Section. In Section III, the proposed ULGFBP-ResNet51 method is presented in details. Afterwards, discussion on the experiments and the paper conclusion are explicated in Section IV and Section VI, respectively.

Table 1 The Pros and Cons of some used methods for COVID-19 identification from images

| Method | Advantage | Disadvantage |
|---|---|---|
| LBP [32] | - Computational simplicity<br>- Good performance for grayscale texture | - Creating long histograms<br>- Sensitive to noise, rotation, and illumination<br>- Capturing limited structural information<br>- Increasing the feature size with the neighbours number |
| HOG [33] | Invariance to photometric and geometric changes | Complex computations |
| GF [34] | Robust to features in multi-scale and direction | Complexity in parameter setting |
| CNN [35] | Learning features by removing unimportant parameters | -Need to large raw images<br>-High computational complexity |

## II. RELATED WORK

LBP is one of the most efficient texture descriptors which has been exerted in many feature extraction and processing tasks [16, 36-44]. In [6], they apply the same feature extraction method on X-ray images for COVID-19 identification. Then, the normal, COVID-19, and bacteria pneumonia chest images have been classified using k Nearest Neighbour (k-NN). After 10-fold cross-validation, 96% accuracy had been achieved.

In [45], a set of LBP features were utilized using multilayer perceptron for classification. The optimal performance was achieved by multilayer perceptron with seventy hidden layers and radius equal to six in LBP for covid19 identification from X-ray images. Accuracy was 73.34%. Also, LBP and other texture-based methods such as HOG were employed for analysis of COVID19 and annotation of the X-ray images in [46]. Then, COVID19 and Non-COVID19 images were dissociated by Naïve Bayesian and Random Forest.

In addition, LBP has been combined with other methods. For instance, in [8], the sub-band chest X-ray images were elicited by employing LBP and Dual-Tree Complex Wavelet Transform (DT-CWT). After that CNN was performed for automatic classification between non-COVID-19 and COVID-19 from these images. The best-achieved accuracy was 99.06%.

Li *et al.* [47] have made a multi-task learning framework with an explainable multi-instance learning for multi-lesion segmentation and COVID-19 diagnosis. Their method [47] can learn task-related features adaptively with learnable weights. However, this method has experimented with only CT images.

Yasar and Ceylan have used a twenty-three-layer CNN architecture in the COVID-19 lung CT images that LBP was applied to them. The highest value of accuracy obtained with the help of lungs CT was 95.32 % [48]. Although LBP can produce fair results, it has drawbacks that affect results. There are some methods that can tackle its problems. One of them is ULBP which can cause better results. Another method that makes better the categorization of COVID-19 is GF. According to the results of research in [31], recognizing COVID-19 is 93% using GF and CNN. Against, it is 85% without GF in lung computed tomography scans.

Besides, the selection of the CNN suitable architecture impresses to diagnose COVID-19 performance. The different architectures that have been noticed in COVID-19 researches as follows: VGG [49-51], ResNet [15, 52-67], GoogLeNet/Inception V1 to V3 [56, 68-70], MobileNet [49, 71, 72], AlexNet [58, 73, 74]. Table 2 reports the Pros and Cons of different CNN architectures.

Among the above-mentioned architectures, ResNet is applied more than others [15, 52-62]. It has generated promising results [3, 12, 26]. In fact, ResNet has resolved the accuracy degradation due to an increase in network layers and depth for improving performance [2].

Using pre-trained models is addressed to decrease the low data problem. Recently, pre-trained ResNet with 50, 101, and 152 layers were suggested for the COVID-19 detection by Narin et al. [26]. The highest efficiency got using pre-trained ResNet50. So that 99.7% was COVID-19 detection accuracy on the X‑ray data. The loss value of ResNet50 was lower than other models, too. It is noteworthy that combining methods with the best performance can lead the COVID-19 identification system to proper accuracy.

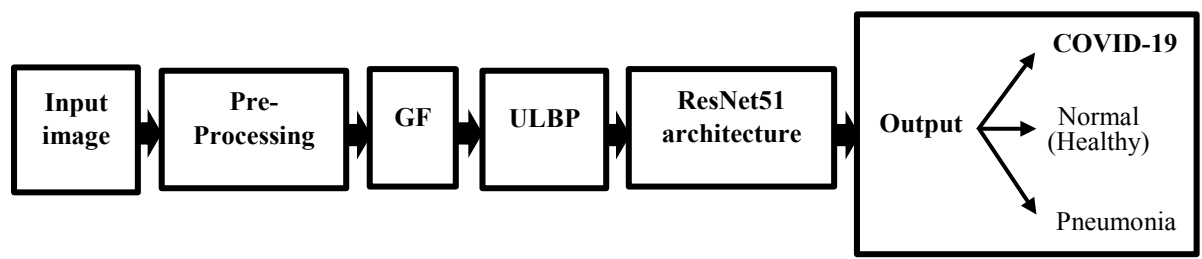

**Fig. 1** The design diagram of the ULGFBP-ResNet51 for COVID-19 diagnosis on lung and chest images

**Table 2** The Pros and Cons of different CNN architectures

| Architecture name | Advantages | Disadvantages |
|---|---|---|
| AlexNet [73] | - Reduction of the images classification error rate to half<br>- Low loss of features | - The less depth in comparison to other architectures<br>- Requiring more time to achieve high accuracy |
| VGG [50] | Refining the performance by gaining depth. | -The vanish/detonate gradients by more increasing of the network depth<br>- Slow processing than the ResNet |
| GoogLeNet [69] | -Increasing the network's width and depth<br>- Quicker training than VGG<br>- Reducing the parameters | -High calculation complexity<br>- Modifying the number of channels difficulty |
| MobileNet [71] | -Less number of parameters and weights than GoogleNet<br>- Decreasing the number of network calculations<br>- A lightweight deep CNN<br>- Reducing size of network<br>- low-latency | Reduction of the accuracy slightly |
| ResNet [52] | - Solving the degradation problem<br>- Solving the vanishing gradient problem<br>- Reducing the training time<br>- Requiring a feature learning only once | Increasing architecture complexity |

## III. PROPOSED METHOD

To accommodate all aforementioned benefits for COVID-19 diagnosis, we combine GF, ULBP, and ResNet51 that has a new Fully-Connected Layer (FCL) more than ResNet50. The proposed method (ULGFBP-ResNet51) will be described in details below.

*A. Pre-processing*

The grayscale images are taken and then resized. Increasing the contrast is done by histogram equalization. Besides, the standard deviation and mean techniques are applied to normalize the images.

*B. GF*

GF performance is similar to human visual perception [75]. SO, it is capable of texture interpretations well [76]. It representations frequency content in specific orientations for image texture analysis. In fact, it can achieve a resolution optimally in frequency and space domains [77, 78].

We extract features of the chest and lung images by GF with different directions and scales. Because GF can make effectively individuate between Normal and COVID-19 from all detailed frequencies texture information [79]. The multiple Gabor Magnitude Image (GMI) is obtained using a bank of multi-scale and multi-direction GFs in the frequency domain. The GF is defined as (1) and the Gabor representation is derived by (2):

$$G(x_p, y_p) = \frac{1}{2\pi\delta^2} e\left[-\frac{x_p^2 + y_p^2}{2\delta^2}\right] \cdot \left[e(j\omega x_p) - e\left(-\frac{\omega^2\delta^2}{2}\right)\right] \tag{1}$$

$$G(x, y) = G(x_p, y_p) * I(x, y) \tag{2}$$

where $x_p = x\cos(\theta) + y\sin(\theta)$, $y_p = -x\cos(\theta) + y\sin(\theta)$. $\delta$ and $\omega$ define the direction and scale of GF. $\delta = \frac{\pi}{\omega}$.
∗ means convolution.

*C. ULBP*

The ULGFBP map is obtained by applying ULBP to GMI (Fig. 2). ULBP is a grayscale operator. It generates 59 bins histogram instead of the 256-bins histogram of LBP when the sampling point number is 8. In other words, the ULBP histogram has just 59 optimized output labels.

Suppose there are eight Sampling Points (*SP*) around a pixel (see part (b) in Fig. 3). The ULBP method compares the Gray Value (*GV*) of that pixel to each of its left-top, left-bottom, right-bottom, left-middle, right-top, right-middle, top, and bottom neighbors. Where the pixel's value is lower than the neighbor's value, code "0". Otherwise, code "1" (see part (c) in Fig. 3). Follow the codes clockwise. It gives an eight-digit binary number that can convert to decimal (see part (d) in Fig. 3). This eight-digit binary number is a uniform pattern when has at most two 0 to1 or 1 to 0 transitions. For example, 01000000 has 2 transitions. Thus, it is uniform.

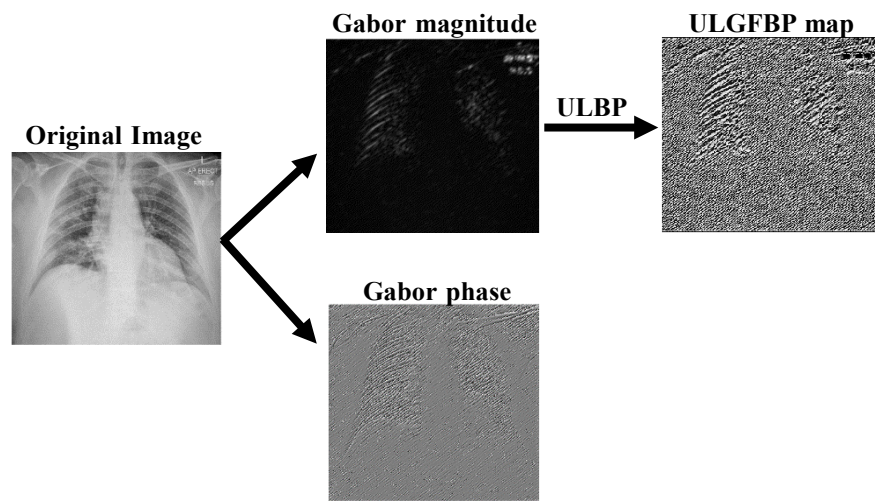

**Fig. 2** The ULGFBP map

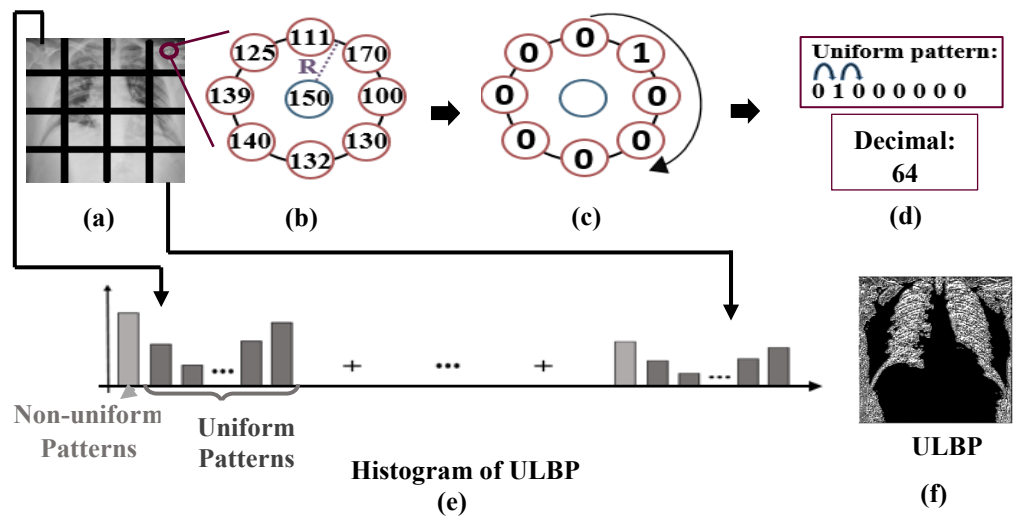

**Fig. 3** The ULBP computation. (a) input image (b) 8 SP around a pixel (c) comparing the central GV with neighbours to set code "0" or "1" (d) convert an eight-digit binary number to decimal (e) concatenating histogram of regions as the ULBP histogram (f) the ULBP

In the computation of the ULBP histogram for each region, a bin is assigned to all non-uniform patterns, and every uniform pattern puts into a separate bin. The ULBP histogram concatenates regions histogram as the global feature vector (see part (e) in Fig. 3). The short feature vector and never changing on rotation are the main advantages of this method [38, 39]. Mathematically, ULBP and its histogram are:

$$ULBP_{SP,R} = |sign(GV_{SP} - GV_c) - sign(GV_0 - GV_c)| + \sum_{es=1}^{SP-1}|sign(GV_{ep} - GV_c) - sign(GV_{ep-1} - GV_c)| \quad (3)$$

$$HULBP = \begin{cases} \sum_{ep=0}^{SP-1} sign(GV_{ep} - GV_c) & if\ ULBP_{SP,R} \leq 2 \\ SP + 1 & otherwise \end{cases} \quad (4)$$

where *ep* represents each sampling point. $R$ is the neighborhood radius where is shown in part (b) of Fig. 3. $c$ means middle pixel.

*D. ResNet51*

ResNet51 is an abbreviation of a 51-layer Residual neural Network. Fifty-one layers are suit due to the time complexity of increasing more the network layers. The bottleneck design can reduce this complexity, too. The ResNet50 model as an improved version of CNN is trained on the ImageNet dataset. In fact, it is trained with about fourteen million various images [26, 80]. Although its network is more complex and deeper, distortion prevention is its profit. In addition, it has fast training due to bottleneck blocks [26].

The ResNet51 is proposed for efficient COVID−19 identification on the X−ray and lung CT images. In fact, we suggest pre-trained ResNet51that has a new layer more than ResNet50 to extract deep features. Because it can be reused with smaller and similar datasets. The model architecture is displayed in Fig. 4. The optimization of the proposed model is easy and it also can produce high accuracy. Its skip connections can solve the vanishing gradient problem.

The input of the network is the ULGFBP map instead of the original image to reach efficient performance. Therefore, the network can extract deep, essential, and robust features. We create a new FCL and also alter the last pre-trained ResNet50 layers according to our data. In fact, we adapt them to our categorization task. These layers are softmax, fully connected, and classification layers. In another word, we replace the last three layers with new layers. Therefore, ResNet51 is made. Output size represents the three COVID-19, pneumonia, and normal classes.

In summary, the ULGFBP maps are passed through the modified pre-trained ResNet51 to obtain features and classified them to COVID-19, pneumonia, and normal using the network classifier. The input size is 224×224×3. Fig. 4 demonstrates the framework of the ULGFBP-ResNet5.

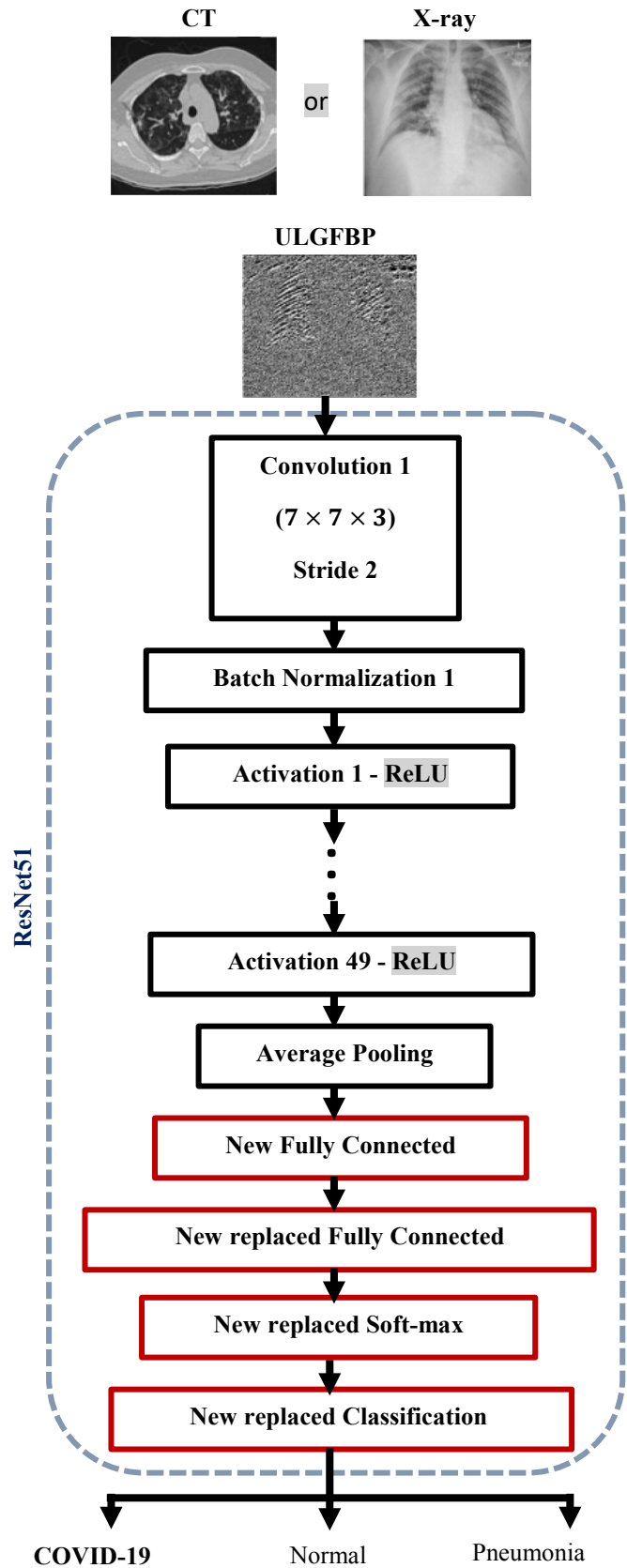

**Fig. 4** The framework of the ULGFBP-ResNet51 method

IV. EXPERIMENTS AND RESULTS

We analyze the implementation of the ULGFBP-ResNet51 in this section. Also, we express its fruitfulness on used datasets. Experiments are done using MATLAB 2020 using a processor of Intel Core i7 and RAM of 8GB onboard.

*A. Datasets*

One of the prerequisites in deep learning is huge training images. We use the dataset[1] collected by El-Shafai and Abd El-Samie for COVID−19 [81]. Its X-ray and CT image data consist of COVID-19 and normal. It is gathered from the GitHub Cohen data [82] and further datasets on internet. Augmentation techniques have been applied to generate more than 17000 images. There are 4044 COVID-19 and 5500 normal X-ray data in a folder. In addition, 5427 COVID-19 and 2628 normal CT data are put in another folder.

Besides, we utilize the pneumonia dataset[2] published by Kermany *et al*. [83]. It has 5232 chest X-ray data. 3883 images belong to pneumonia-infected patients, and 1349 images are collected from healthy persons. A sample of normal, COVID-19, and pneumonia X-ray and CT data from used datasets is illustrated in Fig. 5.

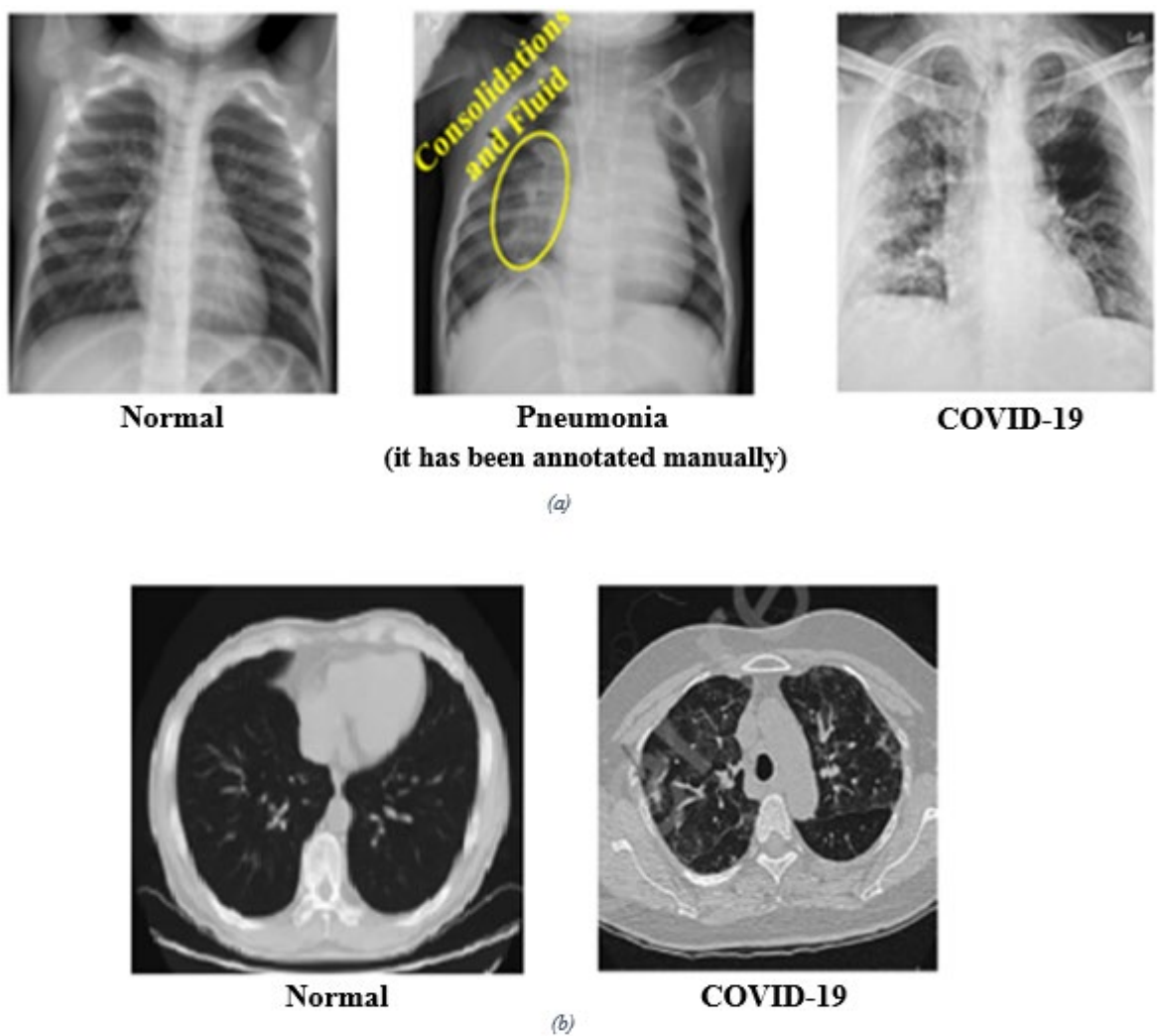

Fig. 5 A sample of the used data. (a) X-ray (b) CT

---

[1] https://data.mendeley.com/datasets/8h65ywd2jr/3
[2] https://data.mendeley.com/datasets/rscbjbr9sj/2

*B. Proposed method*

The grayscale CT or X-ray image is received from the above-mentioned datasets. Using histogram equalization improves the contrast of the image. Also, the image is normalized by the standard deviation and mean techniques.

We extract features using six filters contain two directions and three scales GF. In fact, GMI is obtained by them. Selecting these numbers of filters not only is appropriate but also can decrease the feature dimension.

The obtained GMI is partitioned into 3×3 non-overlapping blocks. Thus, the neighborhood radius ($R$) sets to 1. On the other hand, $SP = 4 \times (2R)$ that if $R$ is 1, then $SP$ is 8. Hence, $ep$ is 0 to SP-1 (i.e., 0 to 7). The cause of this parameter selection is achieving the best performance according to our experiments.

ULBP code and ULBP histogram are computed. So that, each neighborhood pixel ($ep$) value is compared to the c value in each block. Where the pixel's value is lower than the neighbor's value, code "0". Otherwise, code "1". An eight-digit binary number is got when moving clockwise and putting the 0 and 1s. If there are at most two 0 to1 or 1 to 0 transitions then, the pattern is uniform. Thus, a histogram separate bin is assigned to it. Otherwise, it is placed into a single bin. Therefore, the histogram has 59 output labels. For other blocks, ULBP and its histogram are calculated. Finally, the histograms from the first block to the last block are concatenated to get the ULGFBP histogram and ULGFBP map.

The ULGFBP map is the input of pre-trained ResNet51. Its size is converted to 224×224×3. We have 4044 COVID-19, 3883 pneumonia, and 6849 normal ULGFBP maps obtained from X-ray images. In addition, we have 5427 COVID-19, and 2628 normal ULGFBP maps procured from CT images. Then, we use the rotation technique to balance the number of maps.

We save the ULGFBP maps from X-ray images in different folders by COVID-19-X, Normal-X, and Pneumonia-X class name. Also, we keep the ULGFBP maps from CT images in different folders by COVID-19-CT and Normal-CT class name. All images (maps) are tagged.

Pre-trained ResNet-51 is modified according to new data. The training parameters are Batch Size (BS), the number of epochs, Initial Learning Rate (ILR). We utilize Adam optimizer and choose a mini BS of 20. The ILR sets on 0.0001. Max epochs are 5.

Notice that we have created a new FCL. The FCL of the network is used for learning the classification function. Further, we have replaced the last three layers of the network with new FCL, softmax, and classification layers. This work is to adapt the network with our categorization task.

We train the network and optimize hyper-parameters. We train it once with ULGFBP maps from X-ray data for three COVID-19, normal, and pneumonia classes. Training took 12 hours and 43 minutes 18 sec. Once again, the network is trained by two classes COVID-19 and Normal ULGFBP maps from CT images. Training took 6 hours and 37 minutes 20 sec.

The train network function is used for training the ResNet51 model. Also, 10−fold cross−validation strategy is employed. The ULGFBP maps are randomly split into ten folds containing roughly the same proportions of the class labels in each fold. In the ten experiments, nine folds are used for training and one for testing. The average accuracy is reported.

*C. Experimental Results and Discussion*

The achieved accuracy is 99.97% for COVID-19 identification using our proposed method. It demonstrates that our proposed method has outperformance to categorize COVID-19 images. In another word, the evaluation of the effectiveness of the proposed method is its accuracy. The loss and accuracy curves for our proposed method using the X-ray and CT datasets are pictured in Fig. 6. As we have seen, the loss is very low and accuracy is so high on both datasets. Besides, the usefulness of the proposed method is measured using the accuracy, F1-score, sensitivity, and precision metrics [84]. The computation of our method's true predictions in the whole of predictions is named precision. The overall measure of the method's accuracy is shown by F1-score, too.

The confusion matrix not only measures the efficiency of our method on the testing dataset but also shows the classified and miss-classified images correctly. In fact, it specifies the potential of our proposed method for COVID-19, pneumonia, and normal classification. This matrix has True Positive (TP), True Negative (TN), False

Positive (FP), False Negative (FN) expected outcomes. The confusion matrix of our proposed method is illustrated in Fig. 7. The identified anomaly with the correct diagnosis is TP. The number of mistake measures is TN. The classified sample as an anomaly diagnosis is FP. The classified anomalies as ordinary are FN.

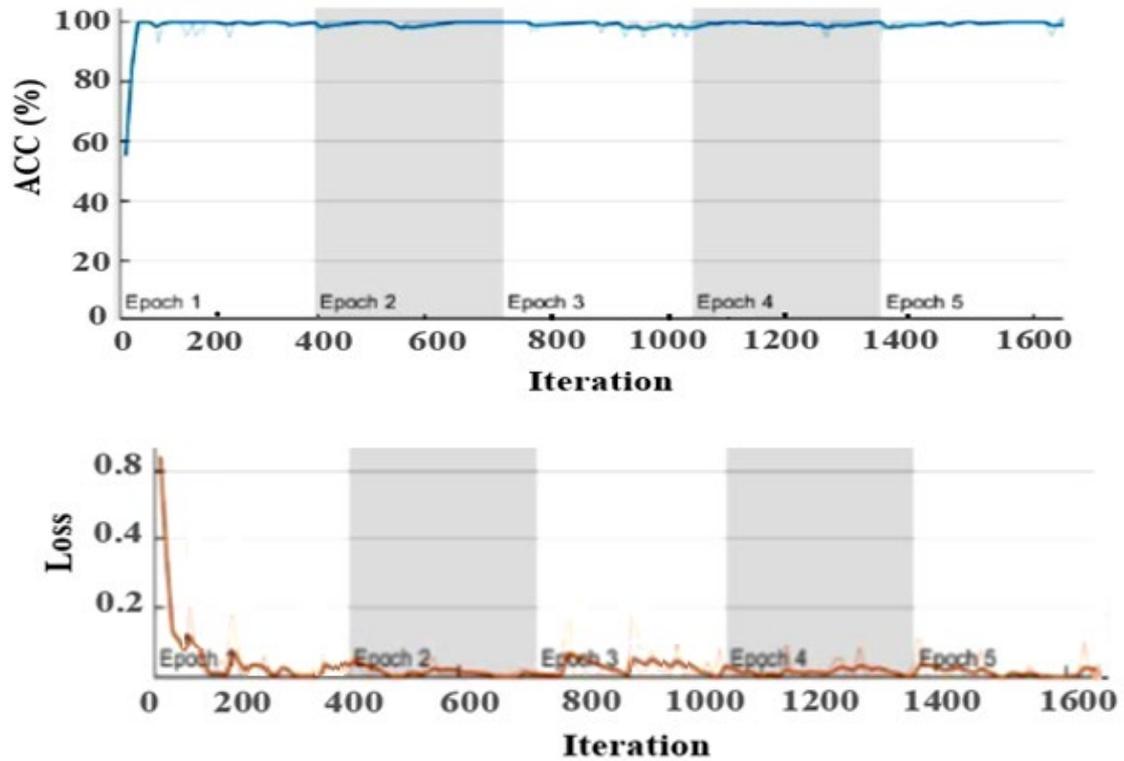

(a)

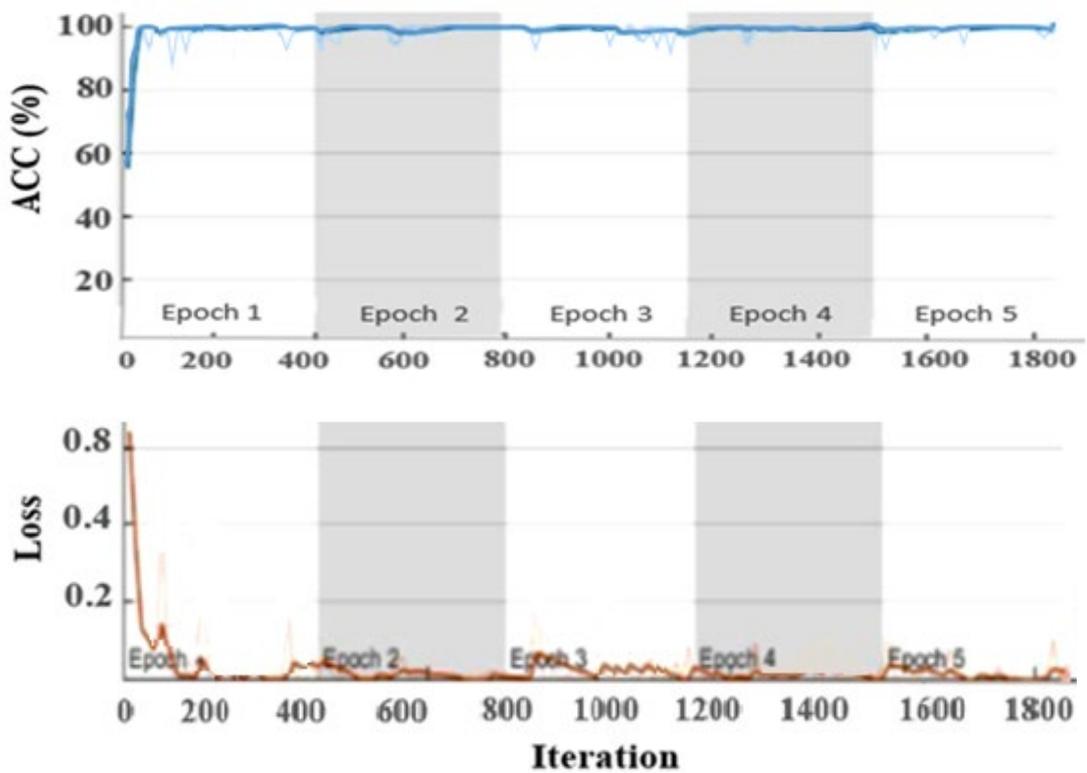

(b)

**Fig. 6** The loss and accuracy curves for our proposed method using datasets (a) CT (b) X-ray

$$Acc\ (\%) = \frac{TN+TP}{TN+TP+FN+FP} \times 100 \tag{5}$$

$$Precision\ (\%) = \frac{TP}{FP+TP} \times 100 \tag{6}$$

$$Sensitivity\ (\%) = \frac{TP}{FN+TP} \times 100 \tag{7}$$

$$F1\ score\ (\%) = 2\left(\frac{Sensitivity \times Precision}{Sensitivity + Precision}\right) \tag{8}$$

Table 4 and Table 5 list the results of the proposed ULGFBP-ResNet51 to compare with other methods. Evaluation of the performance of the ULGFBP-ResNet51 and state-of-the-arts are represented in Fig. 8. It is necessary to point out that due to ResNet51 is 51 layers deep, low time consumption is its advantage in comparison with ResNet101. In addition, its accuracy is superior to other similar models.

## V. CONCLUSION

Economic, education, tourism, transportation, social interaction and on top of that health care system have been hardly devastated by COVID-19. It is proven that by early stage diagnosis of this pandemic, the treatment process proceeds much faster and easier. Therefore, in this manuscript, we propose a novel method called ULGFBP-ResNet51 which could be utilized for autonomous and efficient COVID-19 diagnosis from CT and X-ray datasets. It takes advantage of two texture-based methods (i.e., GF and ULBP) accomplished with deep learning (i.e., ResNet51). Results on COVID-19 datasets demonstrate the outperformance of our approach which provides promising accuracy. The achieved accuracy is 99.97% for COVID-19 images classification. The proposed method could be further exerted on CT and X-ray data in the future for other related diagnostic challenges such as Influenza, tumors and etc.

|  | COVID-19 | Normal | Pneumonia |
|---|---|---|---|
| COVID-19 | 99% | 0% | 1% |
| Normal | 0% | 98% | 2% |
| Pneumonia | 1% | 0% | 99% |

(a)

|  | COVID-19 | Normal |
|---|---|---|
| COVID-19 | 99% | 1% |
| Normal | 3% | 97% |

(b)

**Fig. 7** The confusion matrix of our proposed ULGFBP-ResNet51 method for used datasets (a) X-ray (b) CT

**Table 4** Results of COVID-19 diagnosis using the ULGFBP-ResNet51 and other methods on X-ray

| Method | ACC (%) | Sensitivity (%) |
|---|---|---|
| Contrastive Multi-Task CNN (CMT-CNN) [27] | 97.23 | 92.97 |
| Pre-trained VGG-19 [49] | 93.48 | 92.85 |
| ResNet-50 + SVM [67] | 93.28 | 97 |
| LBP + GF + SVM [18] | 97.16 | 96.9 |
| LBP + DT-CWT + CNN [8] | 99.06 | 96.53 |
| LBP + HOG + SVM [18] | 92.83 | 93.52 |
| ResNet101 [26] | 95.3 | 93.8 |
| *Proposed ULGFBP-ResNet51 method* | *99.97* | *99.9* |

**Table 5** The results of COVID-19 diagnosis using the ULGFBP-ResNet51 and other methods on CT

| Method | ACC (%) | Sensitivity (%) |
|---|---|---|
| Contrastive Multi-Task CNN (CMT-CNN) [27] | 93.46 | 90.57 |
| Random Forest (RF) [19] | 87.9 | 90.7 |
| Deep CNN [28] | 73 | 95 |
| Explainable multi-instance multi-task network [47] | 98.62 | 89.05 |
| *Proposed ULGFBP-ResNet51 method* | *99.9* | *100* |

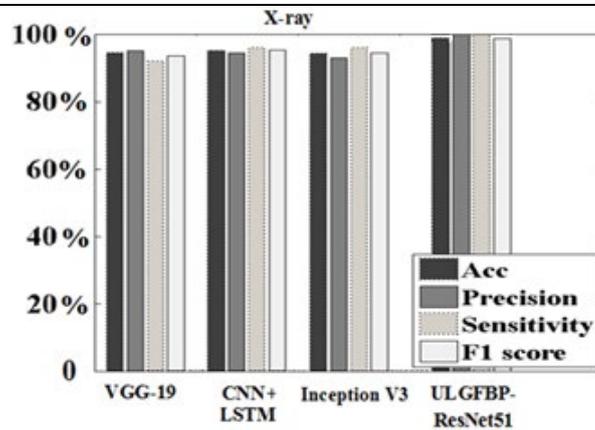

**Fig. 8** Evaluation of the ULGFBP-ResNet51 and state-of-the-arts

**Declarations**

**Funding** This research did not receive any specific grant from funding agencies in the public, commercial, or not-for-profit sectors.

**Conflicts of interest** The authors declare that they have no conflict of interest.

**Availability of data and material** Please contact authors for data requests.

**Code availability** Please contact authors for code request.

**Authors' contributions** All authors took part in the work described in this manuscript.

**Additional declarations for articles in life science journals that report the results of studies involving humans and/or animals**

**Ethics approval (include appropriate approvals or waivers)**   Not applicable.

**Consent to participate (include appropriate statements)**   Not applicable.

**Consent for publication (include appropriate statements)**   Not applicable.

**CRediT authorship contribution statement**

Vida Esmaeili: Conceptualization, Methodology, Investigation, Visualization, Analysis and interpretation, Data curation, Writing - original draft. Mahmood Mohassel Feghhi: Methodology, Investigation, Writing - review & editing, Supervision, Data curation, Validation, Project administration, Formal analysis. Seyed Omid Shahdi: Conceptualization, Methodology, Investigation, Data curation, Visualization, Supervision, Validation, Analysis and interpretation, Writing - review & editing.

REFERENCES

[1] S. Rajpal, N. Lakhyani, A. K. Singh, R. Kohli, N. J. C. Kumar, Solitons, and Fractals, "Using handpicked features in conjunction with ResNet-50 for improved detection of COVID-19 from chest X-ray images," *Chaos, Solitons & Fractals,* vol. 145, p. 110749, 2021.

[2] T. K. Das, C. L. Chowdhary, and X. Gao, "Chest X-Ray Investigation: A Convolutional Neural Network Approach," in Journal of Biomimetics, Biomaterials and Biomedical Engineering, vol. 45, pp. 57-70: Trans Tech Publ, 2020.

[3] M. Ahsan, M. Based, J. Haider, and M. J. S. Kowalski, "COVID-19 detection from chest X-ray images using feature fusion and deep learning," *Sensors*, vol. 21, no. 4, p. 1480, 2021.

[4] H. Alabool *et al.*, "Artificial Intelligence Techniques for Containment COVID-19 Pandemic: A Systematic Review," 2020.

[5] J. R. F. Junior, D. A. C. Cardenas, R. A. Moreno, M. d. F. de Sá Rebelo, J. E. Krieger, and M. A. J. J. o. D. I. Gutierrez, "Novel chest radiographic biomarkers for COVID-19 using radiomic features associated with diagnostics and outcomes," *Journal of Digital Imaging,* pp. 1-11, 2021.

[6] N. Sabri, R. Hamzah, S. Ibrahim, and K. A. F. A. J. I. J. o. E. T. i. E. R. Samah, "COVID-19 detection for chest X-ray images using local binary pattern," *International Journal of Emerging Trends in Engineering Research*, vol. 8, no. 1 Special Issue 1, 2020.

[7] H. Yasar, M. J. M. T. Ceylan, and Applications, "A novel comparative study for detection of Covid-19 on CT lung images using texture analysis, machine learning, and deep learning methods," *Multimedia Tools and Applications*, vol. 80, no. 4, pp. 5423-5447, 2021.

[8] H. Yasar and M. J. A. I. Ceylan, "A new deep learning pipeline to detect Covid-19 on chest X-ray images using local binary pattern, dual tree complex wavelet transform and convolutional neural networks," *Applied Intelligence*, vol. 51, no. 5, pp. 2740-2763, 2021.

[9] Y. Fang *et al.*, "Sensitivity of chest CT for COVID-19: comparison to RT-PCR," *Radiology*, vol. 296, no. 2, pp. E115-E117, 2020.

[10] L. Wang, Z. Q. Lin, and A. J. S. R. Wong, "Covid-net: A tailored deep convolutional neural network design for detection of covid-19 cases from chest x-ray images," *Scientific Reports*, vol. 10, no. 1, pp. 1-12, 2020.

[11] M. N. Kassem and D. T. J. J. o. D. I. Masallat, "Clinical Application of Chest Computed Tomography (CT) in Detection and Characterization of Coronavirus (Covid-19) Pneumonia in Adults," *Journal of Digital Imaging,* pp. 1-11, 2021.

[12] S. Maheshwari, R. R. Sharma, M. J. C. i. B. Kumar, and Medicine, "LBP-based information assisted intelligent system for COVID-19 identification," *Computers in Biology and Medicine*, vol. 134, p. 104453, 2021.

[13] J. E. Luján-García, M. A. Moreno-Ibarra, Y. Villuendas-Rey, and C. J. M. Yáñez-Márquez, "Fast COVID-19 and pneumonia classification using chest X-ray images," *Mathematics,*  vol. 8, no. 9, p. 1423, 2020.

[14] K. Hammoudi *et al.*, "Deep learning on chest x-ray images to detect and evaluate pneumonia cases at the era of covid-19," *Journal of Medical Systems*, vol. 45, no. 7, pp. 1-10, 2021.

[15] O. Albahri *et al.*, "Systematic review of artificial intelligence techniques in the detection and classification of COVID-19 medical images in terms of evaluation and benchmarking: Taxonomy analysis, challenges, future solutions and methodological aspects," *Journal of infection and public health*, 2020.

[16] V. Esmaeili and S. O. Shahdi, "Automatic micro-expression apex spotting using Cubic-LBP," *Multimedia Tools and Applications,* vol. 79, no. 27, pp. 20221-20239, 2020.

[17] Ş. Öztürk, U. Özkaya, M. J. I. J. o. I. S. Barstuğan, and Technology, "Classification of Coronavirus (COVID-19) from X-ray and CT images using shrunken features," *International Journal of Imaging Systems and Technology*, vol. 31, no. 1, pp. 5-15, 2020.

[18] D. Al-Karawi, S. Al-Zaidi, N. Polus, and S. J. m. Jassim, "Ai based chest x-ray (cxr) scan texture analysis algorithm for digital test of covid-19 patients," *medRxiv,* 2020.

[19] F. Shi, L. Xia, and F. Shan, "Large-scale screening of covid-19 from community acquired pneumonia using infection size-aware classification," *arXiv e-prints* [Preprint], 2020.
[20] S. O. Shahdi, and S. Abu-Bakar, "Frequency domain feature-based face recognition technique for different poses and low-resolution conditions," *in 2011 IEEE International Conference on Imaging Systems and Techniques*, pp. 322-326: IEEE, 2011.
[21] V. Esmaeili, M. M. Feghhi, and S. O. Shahdi, "Micro-Expression Recognition Using Histogram of Image Gradient Orientation on Diagonal Planes," *in 2021 5th International Conference on Pattern Recognition and Image Analysis (IPRIA)*, pp. 1-5: IEEE, 2021.
[22] S. O. Shahdi, and S. A. Abu-Bakar, "Variant pose face recognition using discrete wavelet transform and linear regression," *International Journal of Pattern Recognition and Artificial Intelligence*, vol. 26, no. 06, p. 1256013, 2012.
[23] S. O. Shahdi, and S. A. Abu-Bakar, "Varying pose face recognition using combination of discrete cosine & wavelet transforms," *in 2012 4th International Conference on Intelligent and Advanced Systems (ICIAS2012)*, vol. 2, pp. 642-647: IEEE, 2012.
[24] E. Irmak, "COVID-19 disease severity assessment using CNN model," *IET Image Processing*, vol. 15, no. 8, pp. 1814-1824, 2021.
[25] M. Farooq and A. Hafeez, "Covid-resnet: A deep learning framework for screening of covid19 from radiographs," *arXiv preprint arXiv:2003.14395*, 2020.
[26] A. Narin, C. Kaya, and Z. Pamuk, "Automatic detection of coronavirus disease (covid-19) using x-ray images and deep convolutional neural networks," *Pattern Analysis and Applications,* vol. 24, no. 3, pp. 1207-1220, 2021.
[27] J. Li et al., "Multi-task contrastive learning for automatic CT and X-ray diagnosis of COVID-19," *Pattern Recognition*, vol. 114, p. 107848, 2021.
[28] Q. Ni et al., "A deep learning approach to characterize 2019 coronavirus disease (COVID-19) pneumonia in chest CT images," *European radiology*, vol. 30, no. 12, pp. 6517-6527, 2020.
[29] C. Hu, D. Jiang, H. Zou, X. Zuo, and Y. Shu, "Multi-task micro-expression recognition combining deep and handcrafted features," *in 2018 24th International Conference on Pattern Recognition (ICPR)*, pp. 946-951: IEEE, 2018.
[30] M. J. Horry, S. Chakraborty, B. Pradhan, M. Fallahpoor, C. Hossein, and M. Paul, "Systematic investigation into generalization of COVID-19 CT deep learning models with Gabor ensemble for lung involvement scoring," *European radiology*, vol. 30, no. 12, pp. 6517-6527, 2021.
[31] B. V. Krishna, P. N. R. Bodavarapu, P. Santhosh, and P. Srinivas, "Chest Computed Tomography Scan Images for Classification of Coronavirus by Enhanced Convolutional Neural Network and Gabor Filter," *in 2021 5th International Conference on Intelligent Computing and Control Systems (ICICCS)*, pp. 825-831: IEEE, 2021.
[32] T. Ojala, M. Pietikäinen, and D. Harwood, "A comparative study of texture measures with classification based on featured distributions," *Pattern recognition*, vol. 29, no. 1, pp. 51-59, 1996.
[33] N. Dalal, and B. Triggs, "Histograms of oriented gradients for human detection," *in 2005 IEEE computer society conference on computer vision and pattern recognition (CVPR'05)*, vol. 1, pp. 886-893: Ieee, 2005.
[34] D. Gabor, "Theory of communication. Part 1: The analysis of information," *Journal of the Institution of Electrical Engineers-Part III: Radio and Communication Engineering*, vol. 93, no. 26, pp. 429-441, 1946.
[35] Y. LeCun, L. Bottou, Y. Bengio, and P. J. P. o. t. I. Haffner, "Gradient-based learning applied to document recognition," *Proceedings of the IEEE*, vol. 86, no. 11, pp. 2278-2324, 1998.
[36] V. Esmaeili, M. Mohassel Feghhi, and S. O. Shahdi, "Automatic Micro-Expression Apex Frame Spotting using Local Binary Pattern from Six Intersection Planes," *arXiv preprint arXiv: 2104.02149*, 2021.
[37] V. Esmaeili, M. Mohassel Feghhi, and S. O. Shahdi, "Autonomous Apex Detection and Micro-Expression Recognition using Proposed Diagonal Planes," *International Journal of Nonlinear Analysis and Applications*, vol. 11, pp. 483-497, 2020.
[38] S. O. Shahdi, and S. A. R. Abu-Bakar, "Neural network-based approach for face recognition across varying pose," *International Journal of Pattern Recognition and Artificial Intelligence*, vol. 29, no. 08, p. 1556015, 2015.
[39] S. M. Mousavi, S. O. Shahdi, and S. A. R. Abu-Bakar, "Crowd estimation using histogram model classificationbased on improved uniform local binary pattern," *International Journal of Computer and Electrical Engineering*, vol. 4, no. 3, p. 256, 2012.
[40] K. Shankar, and E. Perumal, "A novel hand-crafted with deep learning features based fusion model for COVID-19 diagnosis and classification using chest X-ray images," *Complex & Intelligent Systems*, vol. 7, no. 3, pp. 1-17, 2020.
[41] V. Esmaeili, M. Mohassel Feghhi, and S. O. Shahdi, "A comprehensive survey on facial micro-expression: approaches and databases," *Multimedia Tools and Applications*, 2022.
[42] V. Esmaeili, and M. Mohassel Feghhi, "Diagnosis of Covid-19 Disease by Combining Hand-crafted and Deep-learning Methods on Ultrasound Data," *Journal of Machine Vision and Image Processing*, 2022.
[43] V. Esmaeili, M. M. Feghhi, and S. O. Shahdi, "Early COVID-19 Diagnosis from Lung Ultrasound Images Combining RIULBP-TP and 3D-DenseNet," *In 2022 9th Iranian Joint Congress on Fuzzy and Intelligent Systems (CFIS)*, pp. 1-5: IEEE, 2022.

[44] V. Esmaeili, M. Mohassel Feghhi, and S. O. Shahdi, "Micro-Expression Recognition based on the Multi-Color ULBP and Histogram of Gradient Direction from Six Intersection Planes," *Journal of Iranian Association of Electrical and Electronics Engineers*, 2022.
[45] S. D. Thepade, K. Jadhav, S. Sange, and R. Das, "COVID19 identification from chest X-ray using local binary patterns and multilayer perceptrons," *Journal of Critical Reviews*, vol. 7, no. 19, pp. 4277-4285, 2020.
[46] J. Majumdar, S. Ankalaki, G. Karthik, D. S. Salimath, S. R. Sameer, and S. P. Shetty, "Texture feature for analysis of COVID–19 X-ray images," *Journal of Critical Reviews*, vol. 7, no. 15, pp. 3194-3204, 2020.
[47] M. Li et al., "Explainable multi-instance and multi-task learning for COVID-19 diagnosis and lesion segmentation in CT images," *Knowledge-Based Systems*, vol. 252, p. 109278, 2022.
[48] H. Yaşar and M. Ceylan, "A New Radiomic Study on Lung CT Images of Patients with Covid-19 using LBP and Deep Learning (Convolutional Neural Networks (CNN))," 2020.
[49] I. D. Apostolopoulos, and T. A. Mpesiana, "Covid-19: automatic detection from x-ray images utilizing transfer learning with convolutional neural networks," *Physical and Engineering Sciences in Medicine*, vol. 43, no. 2, pp. 635-640, 2020.
[50] K. Simonyan and A. Zisserman, "Very deep convolutional networks for large-scale image recognition," *arXiv preprint arXiv:1409.1556*, 2014.
[51] S. Vaid, R. Kalantar, and M. Bhandari, "Deep learning COVID-19 detection bias: accuracy through artificial intelligence," *International Orthopaedics*, vol. 44, no. 8, pp. 1539-1542, 2020.
[52] K. He, X. Zhang, S. Ren, and J. Sun, "Deep residual learning for image recognition," i*n Proceedings of the IEEE conference on computer vision and pattern recognition*, pp. 770-778, 2016.
[53] X. Xu et al., "A deep learning system to screen novel coronavirus disease 2019 pneumonia," *Engineering*, vol. 6, no. 10, pp. 1122-1129, 2020.
[54] Y. Song et al., "Deep learning enables accurate diagnosis of novel coronavirus (COVID-19) with CT images," *IEEE/ACM Transactions on Computational Biology and Bioinformatics*, 2021.
[55] O. Gozes et al., "Rapid ai development cycle for the coronavirus (covid-19) pandemic: Initial results for automated detection & patient monitoring using deep learning ct image analysis," *arXiv preprint arXiv:2003.05037*, 2020.
[56] P. K. Sethy, and S. K. Behera, "Detection of coronavirus disease (covid-19) based on deep features," 2020.
[57] S. Jin et al., "AI-assisted CT imaging analysis for COVID-19 screening: Building and deploying a medical AI system in four weeks," *MedRxiv*, 2020.
[58] M. E. Chowdhury et al., "Can AI help in screening viral and COVID-19 pneumonia?," *IEEE Access*, vol. 8, pp. 132665-132676, 2020.
[59] D. Ezzat, and H. A. Ella, "GSA-DenseNet121-COVID-19: a hybrid deep learning architecture for the diagnosis of COVID-19 disease based on gravitational search optimization algorithm," a*rXiv preprint arXiv:2004.05084*, 2020.
[60] S. Chatterjee et al., "Exploration of interpretability techniques for deep covid-19 classification using chest x-ray images," *arXiv preprint arXiv:2006.02570*, 2020.
[61] X. Wu et al., "Deep learning-based multi-view fusion model for screening 2019 novel coronavirus pneumonia: A multicentre study," *European Journal of Radiology*, vol. 128, p. 109041, 2020.
[62] A. Waheed, M. Goyal, D. Gupta, A. Khanna, F. Al-Turjman, and P. R. Pinheiro, "Covidgan: data augmentation using auxiliary classifier gan for improved covid-19 detection," *Ieee Access*, vol. 8, pp. 91916-91923, 2020.
[63] M. Elpeltagy, and H. Sallam, "Automatic prediction of COVID− 19 from chest images using modified ResNet50," *Multimedia Tools and Applications*, vol. 80, no. 17, pp. 26451-26463, 2021.
[64] M. Bolhassani, "Transfer learning approach to Classify the X-ray image that corresponds to corona disease Using ResNet50 pretrained by ChexNet," *arXiv preprint arXiv:2105.08382*, 2021.
[65] S. Walvekar, and D. Shinde, "Detection of COVID-19 from CT images using resnet50," *In 2nd International Conference on Communication & Information Processing (ICCIP),* 2020.
[66] E. E.-D. Hemdan, M. A. Shouman, and M. E. Karar, "Covidx-net: A framework of deep learning classifiers to diagnose covid-19 in x-ray images," *arXiv preprint arXiv:2003.11055*, 2020.
[67] P. Sethy, and S. Behera, "Detection of corona virus Disease (COVID-19) based on Deep Features. Preprints: 2020030300," ed, 2020.
[68] S. Wang et al., "A deep learning algorithm using CT images to screen for Corona Virus Disease (COVID-19)," *European radiology*, pp. 1-9, 2021.
[69] C. Szegedy et al., "Going deeper with convolutions," *in Proceedings of the IEEE conference on computer vision and pattern recognition*, pp. 1-9, 2015.

[70] S. H. Kassani, P. H. Kassasni, M. J. Wesolowski, K. A. Schneider, and R. Deters, "Automatic detection of coronavirus disease (covid-19) in x-ray and ct images: A machine learning-based approach," *arXiv preprint arXiv:2004.10641*, 2020.
[71] A. G. Howard et al., "Mobilenets: Efficient convolutional neural networks for mobile vision applications," *arXiv preprint arXiv:1704.04861*, 2017.
[72] M. Toğaçar, B. Ergen, and Z. Cömert, "COVID-19 detection using deep learning models to exploit Social Mimic Optimization and structured chest X-ray images using fuzzy color and stacking approaches," *Computers in biology and medicine*, vol. 121, p. 103805, 2020.
[73] A. Krizhevsky, I. Sutskever, and G. E. Hinton, "Imagenet classification with deep convolutional neural networks," *Advances in neural information processing systems,* vol. 25, pp. 1097-1105, 2012.
[74] H. S. Maghdid, A. T. Asaad, K. Z. Ghafoor, A. S. Sadiq, S. Mirjalili, and M. K. Khan, "Diagnosing COVID-19 pneumonia from X-ray and CT images using deep learning and transfer learning algorithms," *in Multimodal Image Exploitation and Learning 2021*, vol. 11734, p. 117340E: International Society for Optics and Photonics, 2021.
[75] B. A. Olshausen, and D. J. Field, "Emergence of simple-cell receptive field properties by learning a sparse code for natural images," *Nature*, vol. 381, no. 6583, pp. 607-609, 1996.
[76] M. Imani, "Automatic diagnosis of coronavirus (COVID-19) using shape and texture characteristics extracted from X-Ray and CT-Scan images," *Biomedical Signal Processing and Control*, vol. 68, p. 102602, 2021.
[77] A. YA, "Feature Extraction Techniques for Iris Recognition System: A Survey," *International Journal of Innovative Research in Computer Science & Technology (IJIRCST)*, pp. 2347-5552, 2020.
[78] D. Al-Karawi, S. Al-Zaidi, N. Polus, and S. J. M. Jassim, "Machine learning analysis of chest CT scan images as a complementary digital test of coronavirus (COVID-19) patients," *MedRxiv*, 2020.
[79] D. Al-Karawi, S. Al-Zaidi, N. Polus, and S. J. M. Jassim, "Machine learning analysis of chest CT scan images as a complementary digital test of coronavirus (COVID-19) patients," *MedRxiv*, 2020.
[80] Z. Wu, C. Shen, and A. Van Den Hengel, "Wider or deeper: Revisiting the resnet model for visual recognition," *Pattern Recognition*, vol. 90, pp. 119-133, 2019.
[81] W. El-Shafai and F. Abd El-Samie, "Extensive COVID-19 X-Ray and CT chest images dataset. Mendeley Data, V3," ed, 2020.
[82] J. P. Cohen, P. Morrison, L. Dao, K. Roth, T. Q. Duong, and M. Ghassemi, "Covid-19 image data collection: Prospective predictions are the future," *arXiv preprint arXiv:2006.11988*, 2020.
[83] D. S. Kermany et al., "Identifying medical diagnoses and treatable diseases by image-based deep learning," *Cell*, vol. 172, no. 5, pp. 1122-1131. e9, 2018.
[84] C. Nicholson, "Evaluation metrics for machine learning—accuracy, precision, recall, and F1 defined," ed: Pathmind. http://pathmind.com/wiki/accuracy-precision-recall-f1, 2019.